# Acoustic semimetal with Weyl points and surface states

Tingting Liu, Shengjie Zheng, Hongqing Dai, Dejie Yu, Baizhan Xia*

State Key Laboratory of Advanced Design and Manufacturing for Vehicle Body, College of Mechanical and Vehicle Engineering, Hunan University, Changsha, Hunan, 410082, China

Two-dimensional (2D) topological edge states, immunizing against defects and disorders, have greatly revolutionized our scientific cognition on propagation and scattering of acoustic waves[1-11]. Recently, the similar states have been predicted in three-dimensional (3D) acoustic systems consisting of chiral coupling resonance cavities[12-14]. However, the direct observation of topological surface propagation, especially in a 3D acoustic system with simple scatterers but not complicatedly coupling resonances, has still not been exploited. Here, we design and fabricate a 3D acoustic semimetal composed of rotationally stacked rods. This semimetal not only produces a linear Weyl degeneracy with a charge of 1 (or −1), but also yields a double Weyl point with a charge of 2 (or −2). In its nontrivial gaps, we discover the surface states associated with these Weyl points and experimentally demonstrate the topologically protected one-way propagation of acoustic waves. Due to its good fabricability and topological non-reciprocity, this acoustic semimetal provides an excellent platform for the integration of various acoustic devices, from a macroscopic scale to a nanoscale.

xiabz2013@hnu.edu.cn (Baizhan Xia)

Topological edge states in classical wave systems are usually induced from the Dirac dispersions which are linear in 2D reciprocal spaces[1-11,15-34]. The Dirac conical dispersion is governed by a 2D Dirac Hamiltonian $H(\mathbf{k})=v_x k_x \sigma_x + v_y k_y \sigma_y$, where $v_i$, $k_i$ and $\sigma_i$ ($i=x$ or $y$) are group velocities, momenta and Pauli matrices. In a 3D wave system, the 2D Dirac Hamiltonian is invalid, and replaced by a 3D Weyl Hamiltonian $H(\mathbf{k})=v_x k_x \sigma_x + v_y k_y \sigma_y + v_z k_z \sigma_z$. However, the Weyl point is a more elusive entity than a Dirac point. Dirac conical dispersions are always guaranteed as long as the parity-inversion (P) and time-reversal (T) symmetries are protected. However, such a simple principle is not available for the Weyl degeneracy due to the conflicting requirements between P and T symmetries. With P symmetry, Weyl points at $\mathbf{k}$ and its companion $-\mathbf{k}$ should be opposite, while T symmetry requires them to be same. As a result, Weyl points cannot exist in classical wave systems with PT symmetries.

Weyl point is a source or sink of Berry curvature flux in momentum space. As all degrees of freedom in a Weyl Hamiltonian are exhausted, Weyl points are virtually indestructible, unless two Weyl nodes with opposite charges annihilate each other. The artificial Weyl points and the associated surface propagation were firstly realized in photonic crystals with double gyroid structures[35,36]. Subsequently, Weyl points were also discovered in other photonic systems with arsenide[37,38], phosphide[39,40], woodpile[41,42] and saddle-shaped metallic[43,44] structures. For acoustic waves, a dimerized chain consisted of coupling cavities produced a type-II Weyl point[45]. Recently, Weyl points and Fermi arcs were elaborately observed in 3D chiral phononic systems acting as the synthetic gauge flux[12-14]. The chiral interlayer coupling of this synthetic gauge flux was characterized by waveguides connecting the lower and upper resonance cavities.

Here, we design an acoustic semimetal whose chirality is defined by rotationally stacked rods. We note that similar stacked structures have been used for manipulating photons in 3D optical systems[41,46,47]. Rods are easily fabricated, so this acoustic semimetal can be implemented in a widespread phononic spectrum, from a macroscopic scale to a nanoscale, which is difficult for phononic systems based on complicatedly interlayer coupling resonances. Furthermore, the double Weyl point is discovered in an acoustic semimetal for the first time. Differing from the single Weyl point with linear dispersions along three directions[12,13], the double Weyl point expresses a quadratic in-plane dispersion. The topological charge of the double Weyl point is ±2, but not ±1. The surface states related with Weyl points emerge inside the nontrivial gaps, and the topologically protected one-way propagation of acoustic waves is directly observed.

The acoustic semimetal composed of rotationally stacked rods is shown in Figure 1a. The unit cell containing three layers of rods is in purple. The length, width and height of a rod are respectively $a$=30mm, $b$=8mm and $c$=7.8mm. The total height of the unit cell is $d$=3$c$=23.4mm. Each rod is stacked one-by-one with a counterclockwise 60° in-plane rotation from the layer below. All three stacked rods construct a chiral pattern twisting up along the $z$-direction as a space group with $3_1$ screw operation. Due to its translational symmetry along the $z$-direction, $k_z$ can be viewed as a quantum parameter characterizing the topological characteristics of this acoustic semimetal. The first Brillouin zone (BZ) is depicted in Figure 1b. The grey hexagon highlights the 2D Brillouin zone for a fixed $k_z$.

Figures 2a, 2b and 2c show the band structures of this acoustic semimetal at $k_z$=0, $k_z$=0.5$\pi$/$d$, and $k_z$=$\pi$/$d$. In the $k_x$-$k_y$ plane with $k_z$=0, the first and second bands degenerate to a point at K node. Meanwhile, the second and third bands degenerate to a point at Γ node. In the $k_x$-$k_y$ plane with $k_z$=0.5$\pi$/$d$, both degenerated points are split, resulting in two complete gaps. When $k_z$ further increases to $\pi$/$d$, the lower two bands will degenerate to a point at A node, and the upper two bands will degenerate to a point at H node. To verify that these degenerated points are single and double Weyl

points respectively, we further calculate their 3D bands at $k_z$=1. Figure 2d shows that bands around H node degenerate to a point with linear dispersions along three directions. These linear dispersions indicate that the degenerated point at H node is a single Weyl point. Figure 2e shows that bands around A node are quadratic along the $k_x$ and $k_y$ directions but linear along the $k_z$ direction. This quadratically degenerated point is a double Weyl point which can be viewed as the superposition of two single Weyl points.

Distribution of topological charges of single and double Weyl points are plotted in Figure 1c. In the $k_x$-$k_y$ plane with $k_z$=0, the topological charges of single Weyl points at K node are +1 (yellow dots), while the topological charge of double Weyl point at Γ node is +2 (blue dot). In the $k_x$-$k_y$ plane with $k_z$=±π/d, the topological charges of single Weyl points at G and H nodes are −1 (green dots), while the topological charges of double Weyl points at B and A nodes are −2 (red dots). Thus, the chiral coupling of this acoustic semimetal successfully guarantees that the topological charge signs of Weyl points on the $k_x$-$k_y$ plane with a fixed $k_z$ are the same, corroborating the existence of an acoustic synthetic gauge flux[48]. When $k_z$≠0 and $k_z$≠±π/d, single and double Weyl points are lifted, leading to three isolated bands with two completed gaps. For the first band, the charge 1 flux sinks and the charge 2 flux sources are respectively located in the $k_x$-$k_y$ plane with $k_z$=0 and $k_z$=±π/d (Seeing Supplementary Figure S1). For the third band, the charge 1 flux sinks and the charge 2 flux sources are respectively located in the $k_x$-$k_y$ planes with $k_z$=±π/d and $k_z$=0 (Seeing Supplementary Figure S1). As these planes with different $k_z$ possess different charges, the Berry flux will pass through the $k_x$-$k_y$ plane with a fixed $k_z$[48]. For the second band, the charge 1 flux sources and the charge 2 flux sinks are simultaneously located in the $k_x$-$k_y$ planes with $k_z$=0 and $k_z$=±π/d (Seeing Supplementary Figure S1), indicating that the Berry flux cannot pass through the $k_x$-$k_y$ plane with a fixed $k_z$. By evaluating the eigenvalues of symmetry operators at high-symmetry points[49], we can obtain that the Chern numbers of the first, second and third bands are −1, 0 and +1 respectively, when $k_z=0.5\pi/d$. The Chern number of a nontrivial gap is the sum of Chern numbers of bands bellow it. Thus, the Chern numbers of both gaps are −1, when $k_z=0.5\pi/d$. If $k_z$ is negative, an opposite Chern number can be obtained. The nonzero values of gaps lead to topologically protected surface states. The Chern number −1(+1), when $k_z$ is positive (negative), indicates the clockwise (counterclockwise) surface state.

A supercell consisted of a row of unit cells along the *y*-direction is shown in Figure 3. The hard boundary condition which can be considered as a trivial system is applied at two ends of this supercell. The periodic condition is applied along the *x*- and *z*- directions. If $k_z$ is treated as an additional quantum parameter, we can obtain the corresponding bulk band structure. As shown in Figure 3a, the gapless surface bands emerge in both nontrivial gaps for $k_z=0.5\pi/d$. Blue bands among them denote the surface states localized at the top boundary of supercell (presented in Figures 3b and 3c), propagating along a negative *x*-direction. On the contrary, red bands denote the surface states localized at the bottom boundary (presented in Figures 3b and 3c), propagating along a positive *x*-direction. For the surface states at $k_z<0$, the opposite propagation can be yielded (seeing Supplementary Figure S2). Namely, the surface states located at the top and bottom boundaries propagate along the positive and negative *x*-directions respectively. Therefore, although P symmetry breaking yields two bulk gaps in which the propagation of acoustic waves is efficiently prevented, the surface states will lead to a topologically protected one-way propagation. When $k_z$ is 0 or $\pm\pi/d$, the bulk bands will touch each other, leading to the close of nontrivial gaps (seeing Supplementary Figure S3).

We construct a square acoustic system whose left, top and right boundaries are rigid. The bottom boundary is surrounded by air, so acoustic waves can perfectly leak out. This acoustic system is periodic along the z-direction. Several unit cells are removed from the left boundary to yield defects. We place a sound source (black star) at the top boundary. In Figures 3d and 3e, the excitation frequencies are 6692Hz and 9200Hz which are inside the first and second nontrivial gaps respectively. The surface waves for $k_z=0.5\pi/d$ propagate along the negative x-direction at the top boundary, turn down at the upper-left corner, steer around defects, and continually move along the negative y-direction without backscattering, and finally leaks out at the lower-left corner. Namely, the acoustic waves for $k_z=0.5\pi/d$ robustly propagate in an anticlockwise direction, immunizing against corners and defects. On the contrary, the acoustic waves for $k_z=-0.5\pi/d$ robustly propagates in a clockwise direction, starting from the black star, rounding the upper-right corner and finally leaking out at the lower-right corner without backward scattering, as shown in Figures S4.

A cuboid configuration, created by stacking acoustic semimetals one by one, is provided schematically in Methods and shown in Figure 4a. The number of layers of stacked rods is 11. Two sequential point sources (red star) are placed at the middle of the lowest two layers, as shown in Figure 4a. To excite the surface waves with a fixed $k_z=0.5\pi/d$, the phase of the point source increases by $\pi/2$ from the lowest layer. The identical frequencies for both point sources are 6692kHz. The measured field distribution of acoustic pressure shown in Figure 4c is in agreement with the theoretical prediction shown in Figure 4b. The transient analysis of the acoustic wave packet at different times is conducted. As shown in Figure 4d and Supplementary Movie S1, the surface waves are excited at the first and second layers at $t=2.9886\times10^{-5}$s, and then transport to the third, fifth, seventh, ninth and eleventh layers at $t=2.6898\times10^{-4}$s, $7.3222\times10^{-4}$s, $1.0012\times10^{-3}$s, $1.3001\times10^{-3}$s and $2.0024\times10^{-3}$s, exhibiting a non-reciprocally counterclockwise propagation without obvious reflection or backscattering. As expected, the frequency 8000kHz which is not within the gap do not excite the similar one-way propagation (Supplementary Figure S5 and Movie S2).

To conclusion, we design an acoustic semimetal with single and double Weyl points, and experimentally verify its nonreciprocal surface states which are topologically protected from coupling to the bulk modes. As the topological chirality and the robustly one-way propagation are induced by simply stacked rods, this acoustic semimetal offers an excellent opportunity for exploiting magical Weyl physics and opens up a feasible platform for wide-ranging acoustic device applications, from the macroscopic scale to the nanoscale. Besides the topological surface state supported by Weyl semimetal, other phenomena associated with Weyl dispersion may be also realized, such as diverging and diminishing cross sections at target wavelengthes[50]. Furthermore, the Berry curvature in the vicinity of Weyl point may provide an additional degree-of-freedom for manipulating the propagation of acoustic waves and may lead to a polarization-dependent Hall effect of sound[51].

**Acknowledgments**

The paper is supported by National Natural Science Foundation of China (No.11402083), National Key Research and Development Program of China (2016YFD0701105).

**Author contributions**

B.X. conceived the idea. T. L., S. Z. and H.D. performed the numerical simulation. T.L. and S. Z. fabricated the samples and carried out the experimental measurements. T.L., B.X. and D.Y. wrote this manuscript. All the authors contributed to discussion of the results and manuscript preparation. B.X. supervised all aspects of this work and managed this project.

**Additional information**

Supplementary information is available in the online version of the paper. Correspondence and requests for materials should be addressed to B.X.

**Competing financial interests**

The authors declare no competing financial interests.

**Methods**

**Experiments.** There are 11 layers in this acoustic semimetal. Each unit cell has three rods with the side length $a$=30mm, wide $b$=8mm and height $c$=7.8mm. The tolerance of these rods is ±0.2mm. Headphones are placed at the middle of the lowest two layers in the front surface (red star in Figure 4a). The front, right and back surfaces are covered by plexiglass planes, confining acoustic waves on surfaces of this sample. A microphone (BSWA MPA421) attached to the tip of a stainless-steel rod is inserted into the sample through holes among rods. The moving steps of this microphone along the $x$-, $y$- and $z$- directions are 30mm, $30 \times \sqrt{3}$mm and 23.8mm respectively. The sound signal measured by this movable microphone is analyzed by LMS SCADAS III.

**Simulations.** All full-wave simulations are carried by a commercial FEM software (COMSOL Multiphysics). Because of the huge acoustic impedance contrast between plexiglass and air, the plexiglass is considered a hard boundary. The density and sound speed of air are $\rho$=1.3kg/m$^3$ and $v$=343 m/s. For the bulk band structures in Figure 2, the periodic boundary conditions are applied along all three directions. For the band structures of a supercell in Figures 3a, S2 and S3, the periodic boundary conditions are imposed along the $x$- and $z$- directions, while a rigid boundary condition is applied along the *y*-direction. In Figures 3d-3g, the black star marks the position of the sound source, the black arrows illustrate the direction of propagation of acoustic wave packet, and the purple rhombuses at the bottom represent the air boundary where the sound wave can efficiently leak out. In Figure 4a, two sequential point sources (red star) are placed at the middle of the lowest two layers. In the Figure 4b and 4d, the excitation frequency is 6692 kHz which is inside the nontrivial gap. Rigid boundaries are imposed on the front surface (with a red star in Figure 4a), the right surface and the back surface of sample.

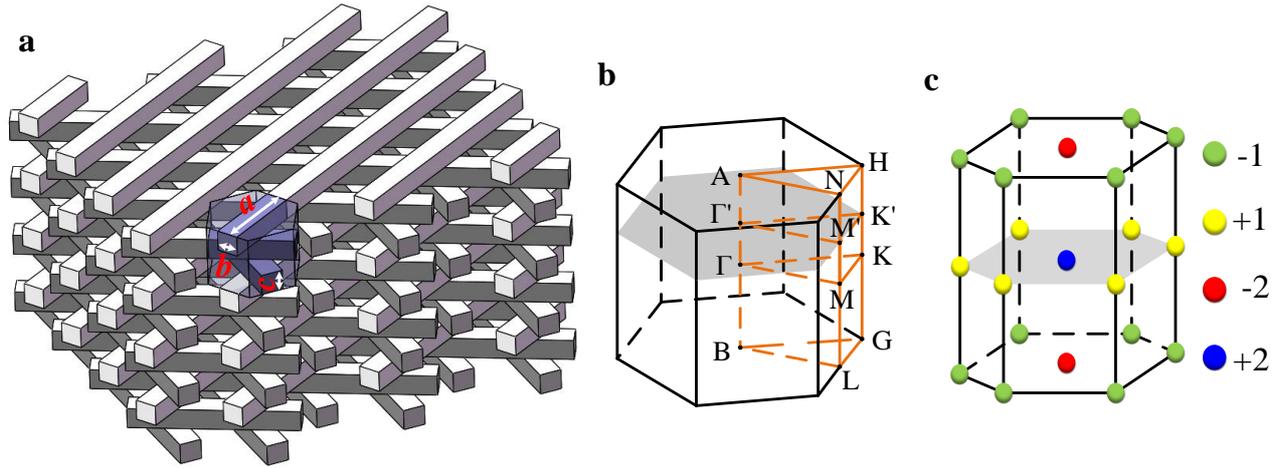

**Figure 1 | Illustration of the acoustic semimetal. a**, Schematic diagram of the acoustic semimetal consisted of rotationally stacked rods. The unit cell contains three layers of rods is in purple. The chiral couplings are introduced by a counterclockwise in-plane rotation of stacked rods. **b**, The first BZ of the acoustic semimetal. The grey hexagon is the first 2D BZ with a fixed $k_z$. **c**, The distribution of topological charges in the 3D reciprocal space. The green and yellow spheres represent the single Weyl points with topological charges -1 and +1 at G (H) and K nodes. The red and blue spheres represent the double Weyl points with topological charges -2 and +2 at B (A) and Γ nodes.

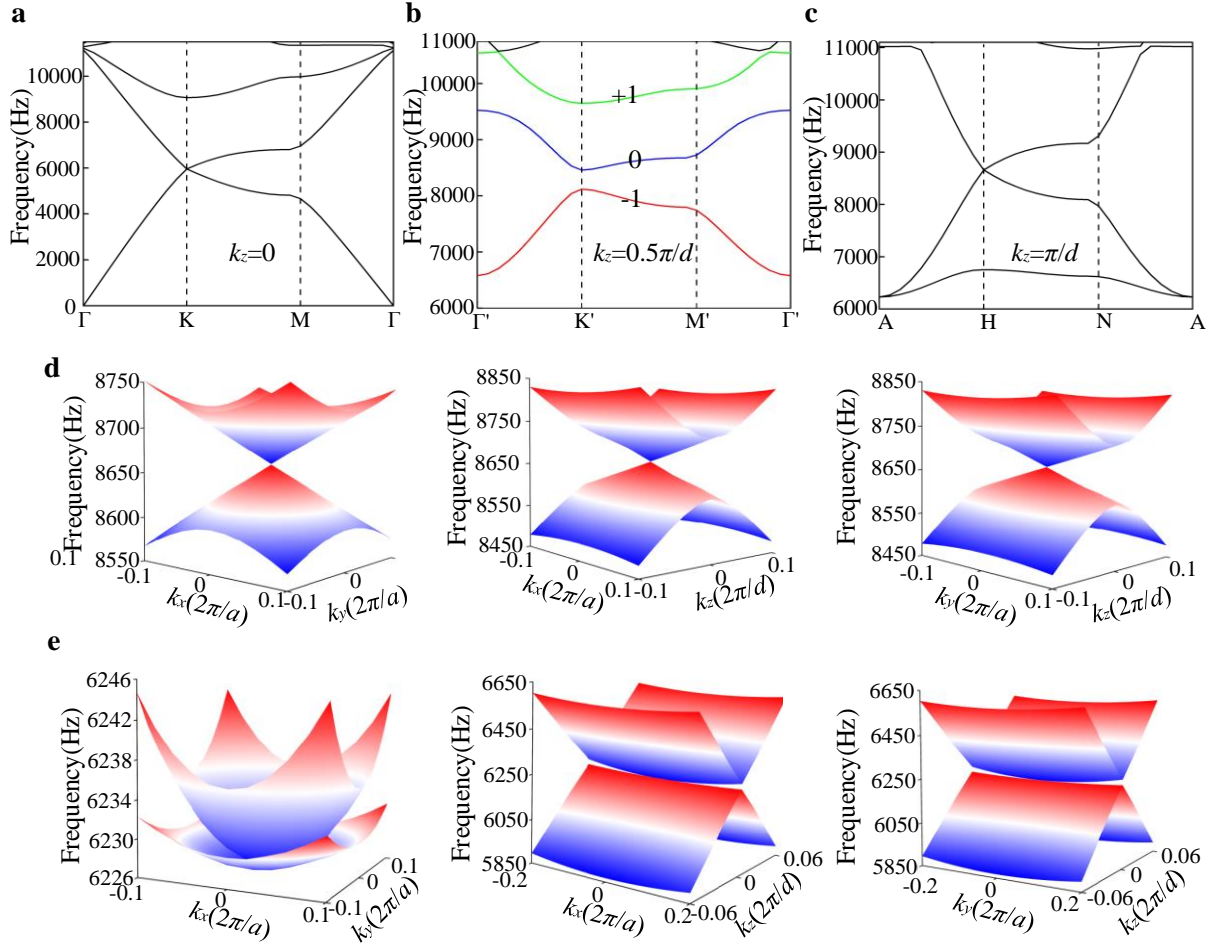

**Figure 2 | Bulk band structures of a unit cell. a**, **b** and **c**, Band structures of the unit cell in the $k_x$-$k_y$ planes at $k_z=0$, $k_z=0.5\pi/d$ and $k_z=\pi/d$. Single Weyl points are found at K and A nodes. Double Weyl points are found at H and Γ nodes. The Chern numbers of three isolated bands are marked in **b**. **d** and **e**, 3D band structures around H and A nodes in the $k_x$-$k_y$ plane (left), the $k_x$-$k_z$ plane (middle) and the $k_y$-$k_z$ plane (right).

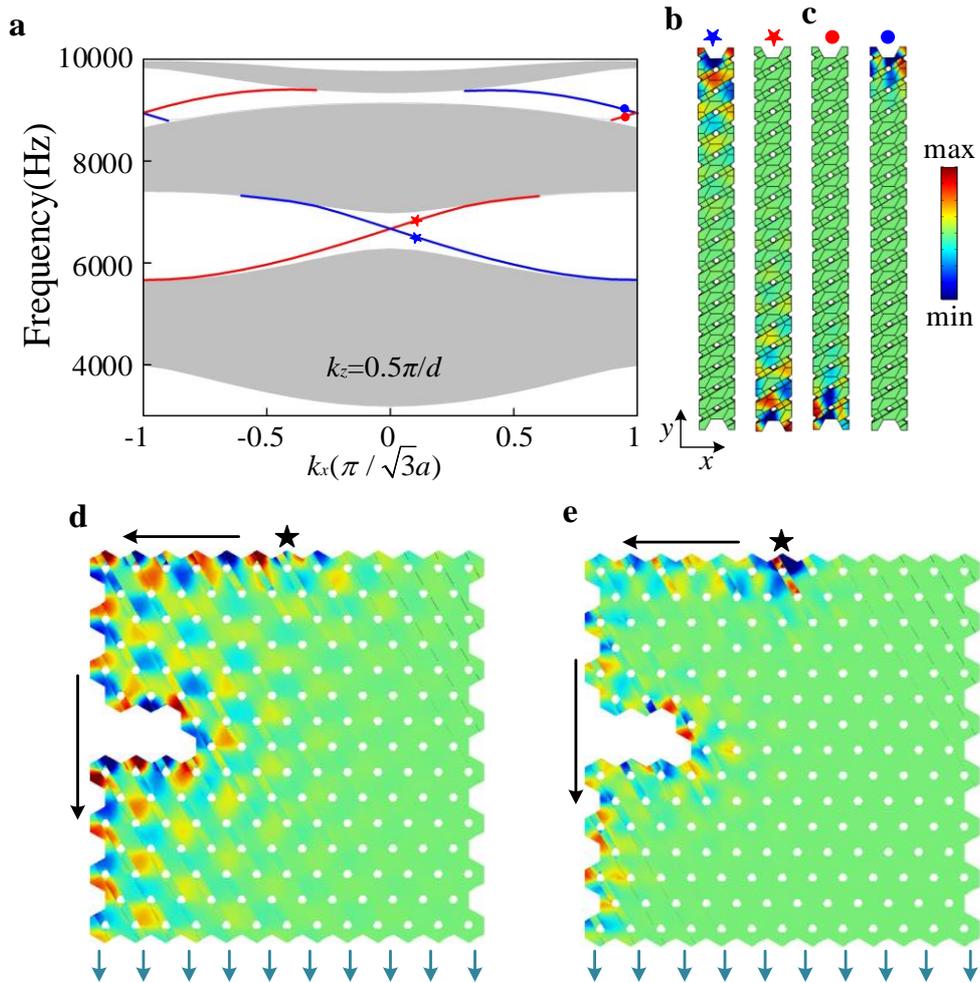

**Figure 3 | Surface sates. a**, Projected band structure of supercell at $k_z=0.5\pi/d$, where blue and red lines denote surface bands. **b**, Eigen pressure fields of surface modes found at $k_x=0.1\pi/d$ marked by blue and red stars in the first nontrivial gap. **c**, Eigen pressure fields of surface modes found at $k_x=0.95\pi/d$ marked by blue and red dots in the second nontrivial gap. **d** and **e**, Counterclockwise propagation of surface waves without backscattering at $k_z=0.5\pi/d$. The excitation frequencies are 6692 kHz (for **d**) and 9200Hz (for **e**) respectively. The black star is the sound source. The black arrow illustrates the direction of propagation. The purple rhombuses at the bottom represent the air boundary where acoustic waves can efficiently leak out.

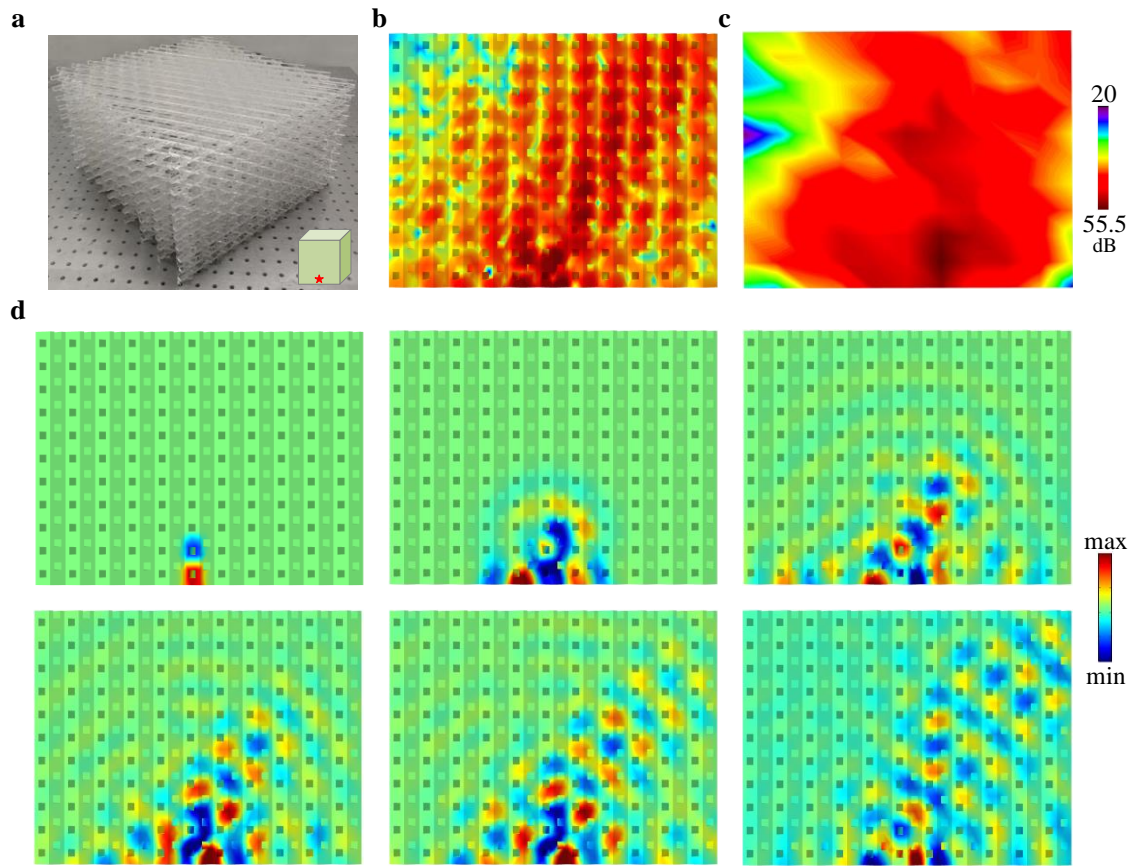

**Figure 4 | Experiment of topologically protected one-way propagation**. **a**, The schematic of a cuboid configuration. **b**, Stimulated field distribution of acoustic pressure at $k_z=0.5\pi/d$. The excitation frequency is 6692 kHz which is inside the first nontrivial gap. **c**, Experimentally measured field distribution of acoustic pressure scanned by a microphone inserted into holes in each layer. **d**, Transient field distributions of acoustic waves on surfaces of the sample at $t=2.9886\times10^{-5}$s, $2.6898\times10^{-4}$s, $7.3222\times10^{-4}$s, $1.0012\times10^{-3}$s, $1.3001\times10^{-3}$s and $2.0024\times10^{-3}$s. Acoustic waves propagate in a counterclockwise direction on surfaces.